\documentstyle[12pt]{article}
\title{Ferromagnetism in the Hubbard model with a~generalized type of hopping}
\author{Pavol Farka\v sovsk\'y\\
Institute  of  Experimental  Physics,  Slovak   Academy   of
Sciences\\
Watsonova 47, 043 53 Ko\v {s}ice, Slovakia}
\date{}
\begin{document}
\baselineskip=24pt
\maketitle

\begin{abstract}

The extrapolation of small-cluster exact-diagonalization
calculations is used to examine ferromagnetism in the one-dimensional 
Hubbard model with a generalized type of hopping. It is found that the 
long-range hopping with power decaying hopping amplitudes 
($t_{ij}\sim q^{|i-j|}$) stabilizes the ferromagnetic state for a wide 
range of electron interactions $U$ and electron concentrations $n$. 
The critical value of the interaction strength $U_c(q)$ above which 
the ferromagnetic state becomes stable is calculated numerically and 
the ground-state phase diagram of the model (in the $U$-$q$ plane) 
is presented for physically the most interesting cases.
\end{abstract}

\newpage
The Hubbard model has become, since its
inception~\cite{Hubbard} in 1963, one of the most popular examples
of a system of interacting electrons with short-range interactions.
It has been used in the literature to study a great variety of many-body
effects in metals, of which ferromagnetism, metal-insulator transitions,
charge-density waves and superconductivity are the most common
examples. Of all these cooperative phenomena the problem of ferromagnetism
in the Hubbard model has the longest history. Although the model was
originally introduced to describe the band ferromagnetism of transition
metals, it soon turned out that the single-band Hubbard model is not
the canonical model for ferromagnetism. In fact the existence of saturated
ferromagnetism has been proven rigorously only for very special
limits. The first well-known example is the Nagaoka ferromagnetism
that comes from the Hubbard model in the limit of infinite repulsion
and one hole in a half-filled band~\cite{Nagaoka}.
Another example where saturated ferromagnetism has been shown to
exist is the case of the one-dimensional Hubbard model with
nearest and next-nearest-neighbor hopping at low electron
densities~\cite{M_H}.
Moreover, several examples of the fully polarized ground state have been
found on special lattices (special conduction bands)
as are the fcc-type lattices~\cite{Ulmke},
the bipartite lattices with sublattices
containing a different number of sites~\cite{Lieb},
the lattices with unconstrained hopping of electrons~\cite{Pieri}
and the flat bands~\cite{M_T}. This indicates that the lattice structure 
and the kinetic energy of electrons, i.e., the type of hopping play 
an important role in stabilizing the ferromagnetic state.  

In this paper we show that if the electron hopping is described 
by a more realistic model (than the nearest-neighbor hopping)
then ferromagnetism comes from the Hubbard model naturally
for a wide range of the model parameters. No extra interactions
terms should be included. In particular, we have found that the
long-range hopping with power decaying hopping amplitudes $t_{ij}$
given by~\cite{Farkas}

\begin{equation}
t_{ij}(q)=\left \{ \begin{array}{ll}
  \quad 0,               &   i=j,\\
  -q^{|i-j|}/q,          &   i\neq j
\end{array}
\right.
\end{equation}
gives rise to ferromagnetism for electron densities above half-filling.
As soon as $q$ that controls the effective range of the hopping
($0\le q \le 1$) is different from zero the ferromagnetic state 
is stabilized for all Coulomb interactions $U$ greater than 
some critical interaction strength $U_c(q)$ that value is 
dramatically reduced with increasing $q$. From this point of view 
one of main reasons why the ferromagnetic state absent in  the ordinary
Hubbard model with nearest-neighbor hopping ($q=0$) is that the description 
of the electron hopping was too simplified.  

The selection of hopping matrix elements in the form given by Eq.~1 has 
several advantages. It represents a much more realistic
type of electron hopping on a lattice (in comparison to
nearest-neighbor hopping), and it allows us to change
continuously the type of hopping (band) from nearest-neighbor $(q=0)$
to  infinite-range $(q=1)$ hopping and thus immediately study the effect 
of the long-range hopping. Another advantage follows from Fig.~1 
where the density of states (DOS) corresponding to Eq.~1 is displayed
for several values of $q$. It is seen that with increasing $q$ more weight
shifts to the upper edge of the band and the DOS becomes strongly asymmetric.
Thus one can study simultaneously (by changing only one parameter $q$)
the influence of the increasing  asymmetry in the DOS
and the influence of the long-range hopping on the ground state properties of
the Hubbard model.

The Hamiltonian of the single-band Hubbard model is given by

\begin{equation}
H=\sum_{ij\sigma}t_{ij}c^+_{i\sigma}c_{j\sigma}+
U\sum_{i}n_{i\uparrow}n_{i\downarrow},
\end{equation}
where $c^+_{i\sigma}$ and $c_{i\sigma}$ are the creation and annihilation
operators  for an electron of spin $\sigma$ at site $i$,
$n_{i\sigma}$ is the corresponding number operator
($N_{\sigma}=\sum_i n_{i\sigma}$) and
$U$ is the on-site Coulomb interaction constant.

The exact results on the ground states of the Hubbard model with 
the generalized type of hopping (Eq.~1) exist only for the special 
case of $q=1$ when the electrons can hop to all sites with equal 
probabilities~\cite{Pieri}. For this type of hopping and the electron 
filling just above half-filling ($N=\sum_{\sigma}N_{\sigma}=L+1$,
where $L$ is the number of lattice sites) it was shown that the ground 
state is not degenerate with respect to the total spin $S$ and it is 
maximally ferromagnetic with $S=(L-1)/2$ (for all $U>0$). For higher 
fillings $(N > L+1)$ the ferromagnetic ground state still exists 
but it is completely degenerate with respect to $S$. The limit of 
infinite-range hopping is, however, the least realistic limit of Eq.~1. 
It is interesting, therefore, to look at the possibility of ferromagnetism 
in the Hubbard model with a generalized type of hopping for smaller values 
of $q$ that describe a much more realistic type of electron hopping.

In this paper we extend calculations to arbitrary $q$ and  
arbitrary band fillings $n=N/L$. The ground states of the model 
are determined by exact diagonalizations for a wide range of model
parameters ($q,U,n$). Typical examples are then chosen from a large 
number of available results to represent the most interesting cases.
The results obtained are presented in the form of phase diagrams 
in the $U$-$q$ plane. To determine the phase diagram in the $U$-$q$
plane (corresponding to some $L$ and $n$) the ground state energy 
of the model is calculated point by point as functions of $q$ and $U$. 
Of course, such a procedure demands in practice a considerable amount 
of CPU time, which imposes severe restrictions on the size of clusters 
that can be studied with this method~($L \sim 16$).
Fortunately, we have found that the ground-state energy of the 
model depends on $L$ only very weakly (for a  wide  range of the 
model parameters) and thus already such small clusters can describe 
satisfactorily the ground state properties of the model.

Let us first briefly discuss the case of $N=L+1$. According to 
analytical results~\cite{Pieri} only for this case the ground state 
of the model with unconstrained hopping $q=1$ is maximally ferromagnetic 
and nondegenerate with respect to the total spin $S$. Numerical 
calculations that we have performed on finite clusters 
up to $L=12$ showed that the ferromagnetic state persists as 
the ground state also for $q<1$, but with decreasing $q$ the 
region of its stability shifts to higher values of $U$. In accordance 
with the exact results~\cite{Pieri} obtained for $q=1$ we found that
the critical interaction strength $U_c(q)$ (above which the ground state
is ferromagnetic) goes to zero for $q\to 1$ while in the opposite limit 
($q\to 0$) $U_c$ tends to infinity. Although the appearance of the
ferromagnetic state at finite $q$ and $N=L+1$ is interesting from the
theoretical point of view, thermodynamically this result is not 
significant if the ferromagnetic state does not persist also for 
higher fillings. Analytical results obtained for $q=1$ predict, however, 
that the ground states for $N > L+1$ are completely degenerate with 
respect to the total spin $S$ and thus the only possibility for the 
stabilization of the ferromagnetic state is that the long-range hopping 
with $q\neq 1$ removes this degeneracy. 

Numerical calculations that we have performed for a wide range of electron
fillings $n>1$ fully confirmed this assumption. It was found that 
the long-range hopping with $q\neq1$ not only removes the degeneracy 
of the ground states with respect to $S$ but at the same time stabilizes 
the ferromagnetic state. Furthermore, these calculations showed that 
the effect of the long-range hopping on the stability of the ferromagnetic 
state is extremely strong, especially for small values of $q$.  
The results of our small-cluster exact-diagonalization calculations
obtained on finite clusters up to $L=16$ sites are summarized in 
Fig.~2. There is shown the critical interaction strength $U_c$, above 
which the ground state is ferromagnetic, as a function of $q$ for several 
values of electron concentrations $n$ ($n=5/4,3/2,7/4$). 
To reveal the finite-size effects on the stability of ferromagnetic domains, 
the behavior of the critical interaction strength $U_c(q)$ has been calculated
on several finite clusters at each electron filling. It is seen that 
finite-size effects on $U_c$ are small and thus these results can be 
satisfactorily extrapolated to the thermodynamic limit $L\to \infty$.           
Our results clearly demonstrate that the ferromagnetic state is
strongly influenced by $q$ for electron concentrations above half-filling
and generally it is stabilized with increasing $q$~\cite{note}. The effect 
is especially strong for small values of $q$ where small changes of 
$q$ reduce dramatically the critical interaction strength $U_c$ and
so the ferromagnetic state becomes stable for a wide range of 
model parameters.  The results presented in Fig.~2d show that only for 
$q=0$ (nearest-neighbor hopping) $U_c=\infty$, while for finite 
$q$ (that represents a much more realistic type of electron 
hopping) the critical interaction strength $U_c$ is finite. Thus the
absence of ferromagnetism in the ordinary Hubbard model with the
nearest-neighbor hopping ($q=0$) can be explained as a consequence of 
too simplified description of electron hopping on the lattice. For 
any $q>0$ ferromagnetism comes naturally from the Hubbard model with 
a generalized type of hopping for a wide range of model parameters 
without any other assumptions.  This opens a new  route towards 
the understanding of ferromagnetism in the Hubbard model.

In order to exclude the possibility that ferromagnetism in the Hubbard 
model with the long-range hopping described above is a consequence of 
boundary conditions used (open boundary conditions), let us finally examine 
the influence of boundary conditions on the stability of the ferromagnetic 
state. In Fig.~3 we present representative results of exhaustive numerical 
studies of the model obtained for $n=3/2$ on all finite (even) clusters up 
to $L=16$ sites for open and periodic boundary conditions. There is plotted 
the difference $\Delta E=E_f-E_g$ between the exact ground state $E_g$ and 
the ferromagnetic state $E_f$ as a function of $1/U$ for $q=0.25$ and 
different $L$. The ground state is ferromagnetic in the regions where 
$\Delta E=0$. In accordance with results discussed above one can
see a nice convergence of numerical results for the open boundary 
conditions. These results can be satisfactorily extrapolated to the 
thermodynamic limit $L=\infty$. We have plotted this extrapolated behavior 
in Fig.~3b to demonstrate clearly a convergence of small-cluster
exact-diagonalization results for the periodic boundary conditions.
It is seen that results for the periodic boundary conditions converge
slowly than ones for the open boundary conditions, but their 
convergence to the thermodynamic limit is apparent. This confirms
the stability of the ferromagnetic state in the Hubbard model with the 
long-range hopping.     

In summary, the extrapolation of small-cluster exact-diagonalization
calculations was used to examine ferromagnetism in the one-dimensional 
Hubbard model with a generalized type of hopping. It was found that 
the long-range hopping with power decaying hopping amplitudes 
stabilizes the ferromagnetic state for a wide range of model 
parameters. The critical value of the interaction strength $U_c(q)$ 
above which the ferromagnetic state becomes stable was calculated numerically 
and the ground-state phase diagram of the model (in the $U$-$q$ plane) was 
presented for selected values of electron fillings.

\vspace{0.5cm}
This work was supported by the Slovak Grant Agency VEGA
under grant No. 2/7021/20. Numerical results were obtained using
computational resources of the Computing Centre of the Slovak 
Academy of Sciences.

\newpage
Figure Captions

\vspace{0.5cm}
Fig.~1. The  non-interacting DOS corresponding to a generalized hopping
(Eq.~1) for different values of $q$ and $L=4000$ sites.

\vspace{0.5cm}
Fig. 2. The critical interaction strength $U_{c}$ ($1/U_c$) as a function 
of $q$ calculated for different $n$ and $L$. 

\vspace{0.5cm}
Fig. 3. The difference $\Delta E=E_g-E_f$ between the exact ground state $E_g$ 
and the ferromagnetic state $E_f$ as a function of $1/U$ calculated for 
$n=3/2, q=0.25$ and different $L$. (a) Open boundary conditions.
(b) Periodic boundary conditions.


\begin{thebibliography}{99}
\bibitem{note} In accordance with the analytical results (Ref.~6)
we have found that the long-range hopping does not stabilize 
the ferromagnetic state for $n\le 1$.
\end{thebibliography}
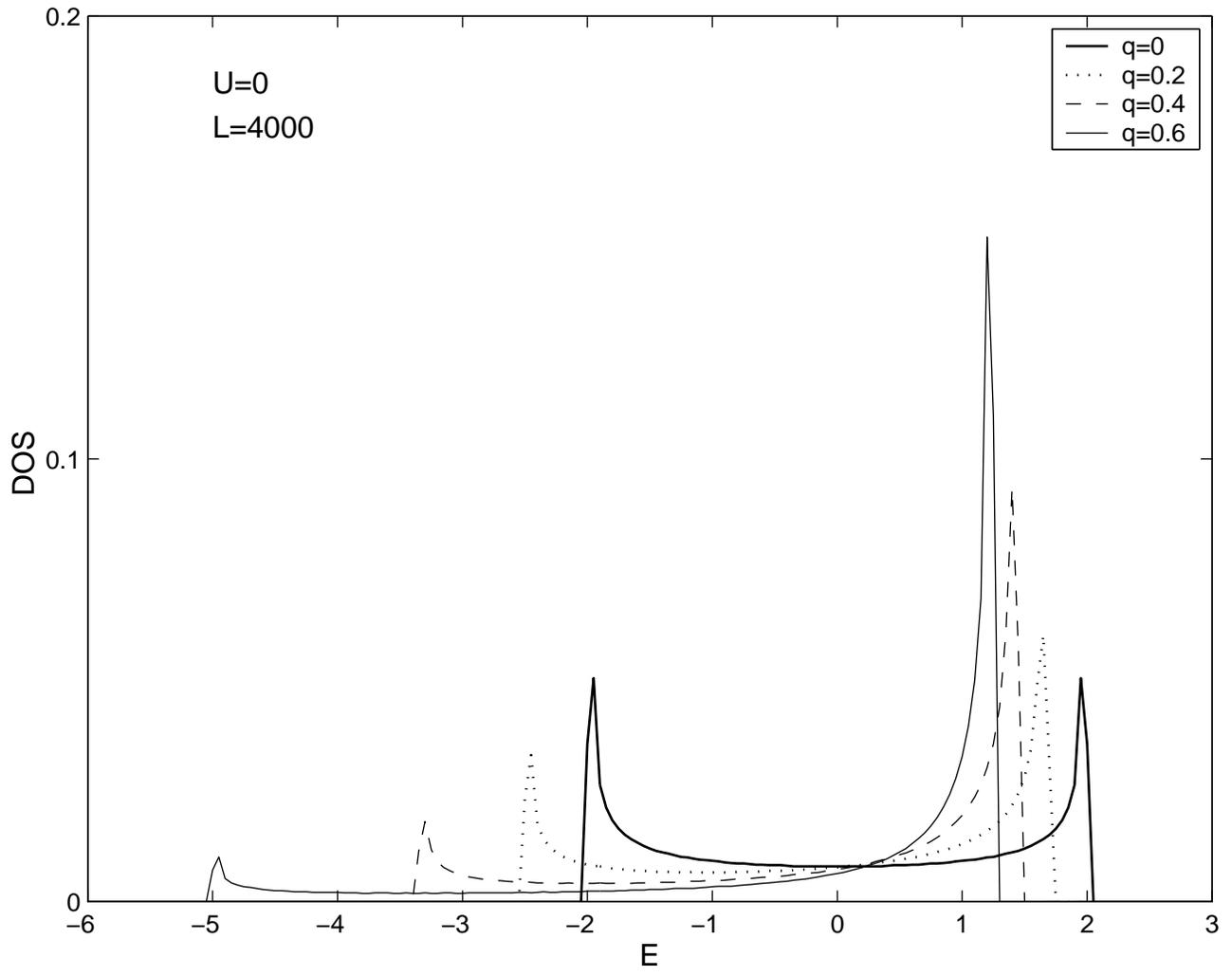
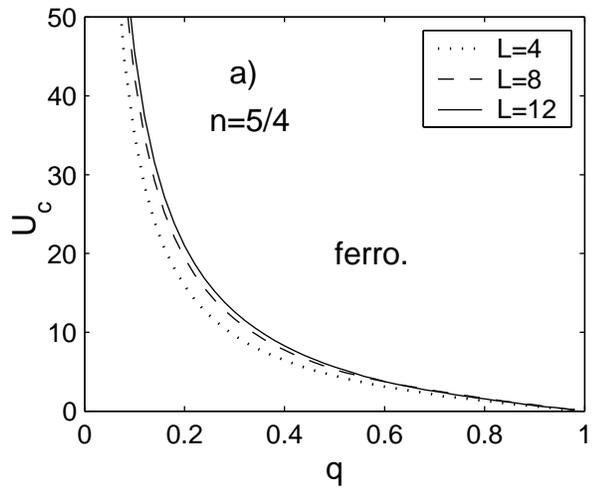
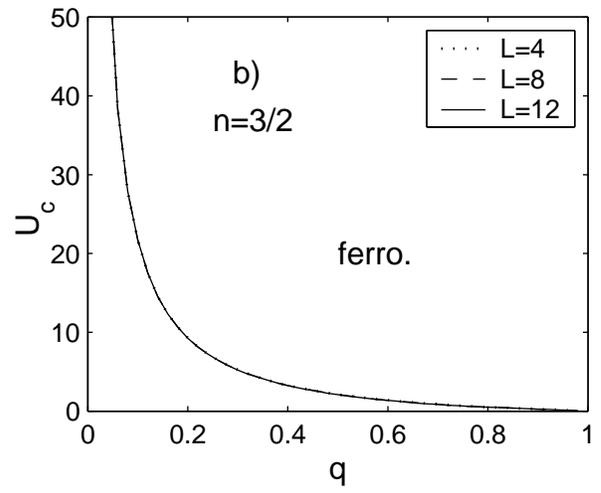
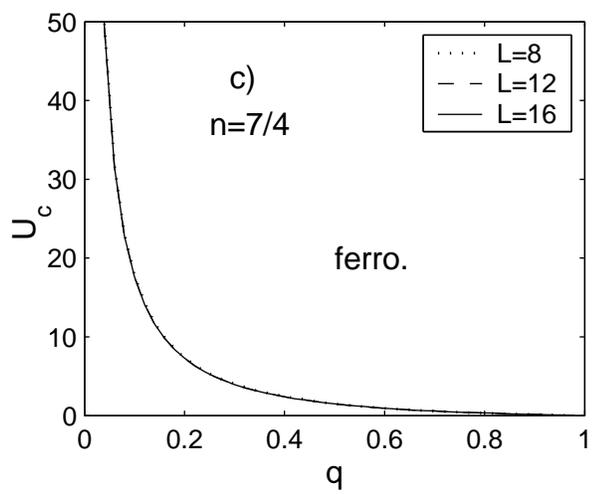
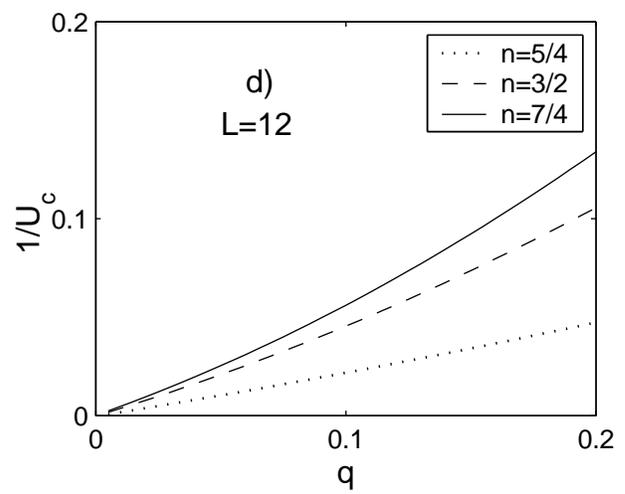
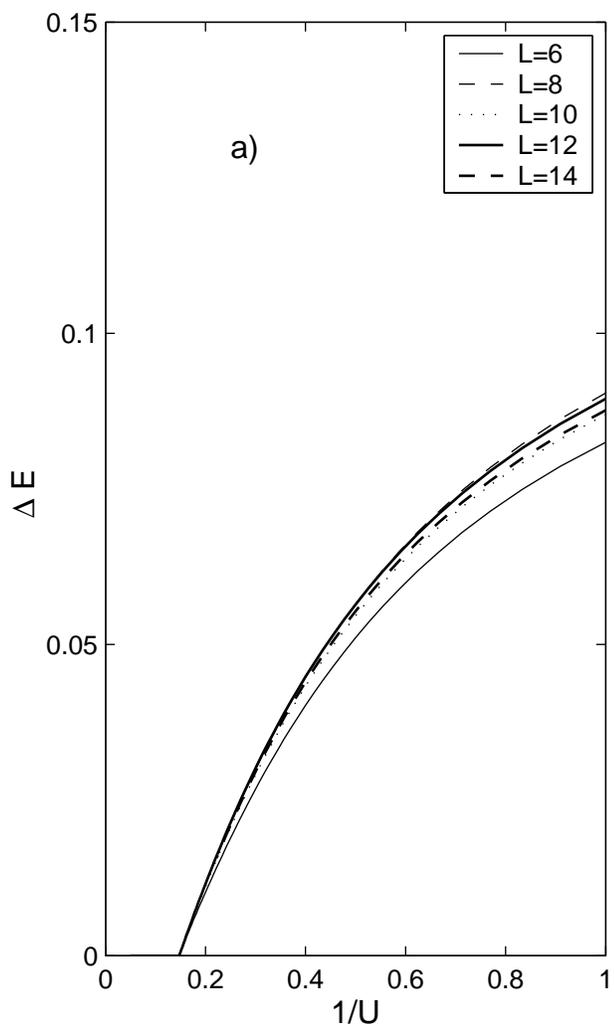
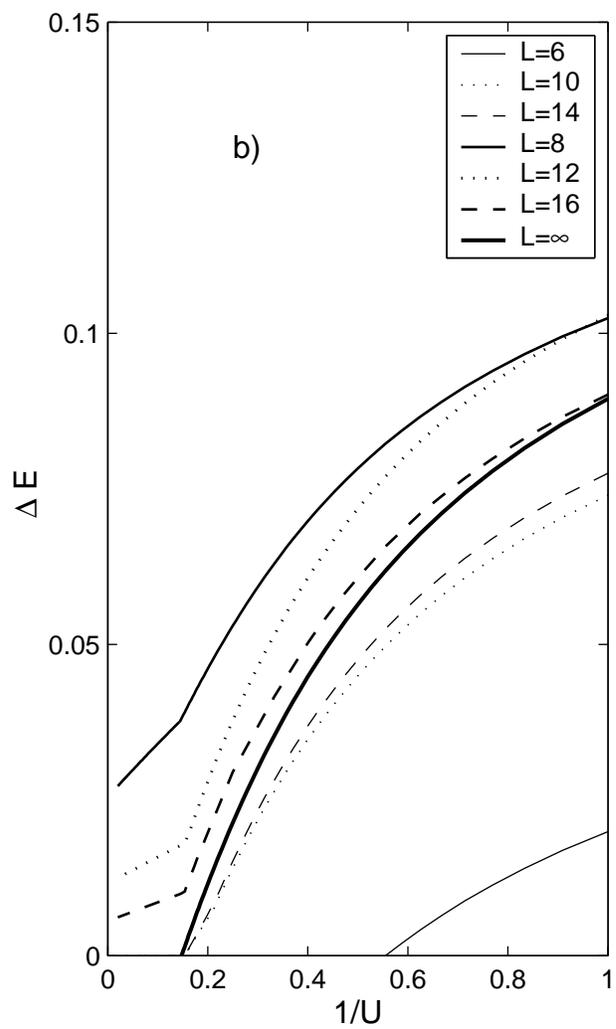
\end{document}